\documentclass[preprint, tightenlines, superscriptaddress, nofootinbib, groupedaddress]{revtex4-1}
\usepackage{amsmath}
\usepackage{amsfonts}
\usepackage{amssymb}
\usepackage{float}
\usepackage{graphicx}
\usepackage{calc}
\usepackage{color}
\usepackage{paralist}
\usepackage{enumitem}
\usepackage[normalem]{ulem}
\usepackage[section]{placeins}

\newcommand{\sub}[1]{\ensuremath{_{\textrm{#1}}}} 
\newcommand{\super}[1]{\ensuremath{^{\textrm{#1}}}} 

\begin{document}
\title{Quantum Critical Ballistic Transport in Two-Dimensional Fermi Liquids}

\author{Mani Chandra}
\email{mani@quazartech.com}
\affiliation{Research Division, Quazar Technologies, Sarvapriya Vihar, New Delhi, India, 110016}

\author{Gitansh Kataria}
\affiliation{Research Division, Quazar Technologies, Sarvapriya Vihar, New Delhi, India, 110016}

\author{Deshdeep Sahdev}
\affiliation{Research Division, Quazar Technologies, Sarvapriya Vihar, New Delhi, India, 110016}

\begin{abstract}
Electronic transport in Fermi liquids is usually Ohmic, because of momentum-relaxing scattering due to defects and phonons. These processes can become sufficiently weak in two-dimensional materials, giving rise to either ballistic or hydrodynamic transport, depending on the strength of electron-electron scattering. We show that the ballistic regime is a quantum critical point (QCP) on the regime boundary separating Ohmic and hydrodynamic transport. The QCP corresponds to a \emph{free} conformal field theory (CFT) with a dynamical scaling exponent $z = 1$. Its nontrivial aspects emerge in device geometries with shear, wherein the regime has an intrinsic universal dissipation, a nonlocal current-voltage relation, and exhibits the critical scaling of the underlying CFT. The Fermi surface has electron-hole pockets across all angular scales and the current flow has \emph{vortices} at all spatial scales. We image the fluctuations in high-definition and animate their emergence as experimental parameters are tuned to the QCP\footnote{Spatial fluctuations: \url{https://vimeo.com/365020115} , Fermi surface fluctuations: \url{https://vimeo.com/364982637}}. The vortices clearly demonstrate that Pauli exclusion alone can produce collective effects, with low-frequency AC transport mediated by vortex dynamics\footnote{Vortex dynamics and frequency crossover: \url{https://vimeo.com/366725650}}. The scale-invariant spatial structure is much richer than that of an interaction-dominated hydrodynamic regime, which only has a single vortex at the device scale.  Our findings provide a theoretical framework for both interaction-free and interaction-dominated non-Ohmic transport in two-dimensional materials, as seen in several contemporary experiments.
\end{abstract}

\maketitle

Quantum critical points (QCP) mediate second-order quantum phase transitions (QPT), produced by tuning non-thermal parameters such as doping, magnetic field or pressure. Systems at QCPs obey universal scaling, and are dominated by \emph{quantum fluctuations}\cite{sachdev-2011}. A prototypical QPT occurs in the 1D quantum Ising model in a transverse magnetic field, realized in CoNb$_2$O$_6$ \cite{coldea-2010}. As the external field is increased, the ground state changes from an ordered phase set by exchange interaction to a field-aligned quantum paramagnet. The transition occurs through a QCP with a dynamical scaling exponent $z = 1$, which connects the correlation length $\xi$ and time $\tau$ via $\tau \propto \xi^z$. Remarkably, the ubiquitous Fermi liquid hosts exactly such a QCP in the form of \emph{free fermions}; a Fermi gas. Absent all interactions, gapless quasiparticles on the Fermi surface obey a relativistic \emph{free-field} conformal field theory (CFT), with the speed of light set by the Fermi velocity $v\sub{F}$ \cite{sachdev-book}. This simple CFT exhibits critical scaling with $z = 1$. Microscopic interactions are indeed irrelevant, as expected at criticality, because there are none.

Although a Fermi gas obeys critical scaling \cite{sachdev-1995}, it is never pictured as a QCP. The single-particle Green's function has a simple pole, not a branch cut characteristic of interacting QCPs \cite{sondhi-1997}; it is unclear how and where critical fluctuations manifest themselves. Further, dissipation is seen as arising from external non-universal factors that relax momentum (e.g., defects, phonons). We show that free fermion transport in \emph{two}-dimensions, realized by a shear, has the \emph{full} complexity expected of a QCP. The angular dispersion of velocities provided by the Fermi surface gives rise to an \emph{intrinsic} universal dissipation set only by the Fermi wavenumber $k\sub{F}$. Scale-invariant fluctuations permeate \emph{all} measurables: the Fermi surface has particle-hole excitations across several angular scales, currents become organized into \emph{vortices} spanning several spatial scales and give rise to tightly correlated spatial density fluctuations. The vortices reveal striking \emph{collective} behavior, belying the single-particle intuition associated with free fermions. Our calculations exploit a new high-resolution computational microscope that is able to resolve the multitude of critical fluctuations.

The emergent dissipation we encounter here has surfaced before repeatedly. In the 1950s, Lindhard\cite{lindhard-1953, lindhard-1954} showed that a perfect conductor has a resistance due to ``zero point'' motion of electrons on the Fermi surface. This resistance appears as the anomalous skin effect \cite{chambers-1950, reuter-1948-2, pippard-1954}. It also appears at low frequencies in the normal phase of $^3$He, and damps the \emph{zero sound} mode through excitation of particle-hole pairs on the Fermi surface \cite{landau-1965, abel-1966}. In classical physics, it causes Landau damping in collisionless plasmas \cite{landau-damping-1965} and ``violent relaxation'' in astrophysical dynamics \cite{lynden-bell-1966}. The underlying mechanism, known as ``phase-mixing'', arises from free-streaming particles with a momentum spread \cite{mouhot-2011, hammett-1992}. The mechanism converts spatial fluctuations into momentum fluctuations, and requires a finite pressure to operate. The thermal pressure in classical systems disappears as $T \rightarrow 0$ and with it, the dissipation. Crucially however, quantum degeneracy pressure in Fermi liquids allows phase-mixing \emph{even} at $T = 0$; a defining property of \emph{quantum} criticality.

We show that the quantum critical free fermion transport is conveniently realized in two-dimensional devices operating in the semiclassical ballistic regime, which occurs at temperatures ($T \gtrsim 20$ K) where quantum phase coherence is negligible, and when quasiparticle scattering in the bulk is sufficiently weak \cite{beenakker-1991}. Quasiparticles then cannot thermalize, and retain memory of their trajectories. This leads to a large number of correlations which spread on a light-cone with speed $v\sub{F}$, similar to those observed in quench studies \cite{calabrese-2006, cheneau-2012}. The regime is established when the correlations envelop the device. For a device of scale $L$ to be correlated, we require a bulk scattering timescale $\tau\sub{bulk}\gtrsim \tau = CL^{z}$ where we obtain the prefactor $C = 1/(2v\sub{F})$ from finite-size scaling. For $L \sim 1$ $\mu$m, we need $\tau\sub{bulk} \gtrsim 0.5$ ps. These requirements have \emph{already} been achieved. Examples include graphene/hBN\cite{mayorov-2011, dean-2013, wang-2013}, and GaAs\cite{brill-1996}.

By tuning bulk scattering, we can induce a \emph{nonequilibrium} transition similar to the second-order QPT in the 1D quantum Ising model. Bulk scattering can either be momentum relaxing (MR) due to electron-phonon, electron-defect and/or Umklapp electron-electron interactions \emph{or} momentum conserving (MC) due to normal electron-electron interactions. Transport in metals is usually dominated by MR scattering which results in a diffusive Ohmic regime. A novel hydrodynamic regime with collective fluid-like behavior, found in graphene\cite{bandurin-2016, krishnakumar-2017, bandurin-2018, berdyugin-2019}, (Ga,Al)As \cite{dejong-1995, molenkamp-1994, braem-2018} and other select materials \cite{gusev-2018, moll-2016, gooth-2018, jaoui-2018}, can arise when MR scattering is weak and MC scattering is strong. We show that the transition from an Ohmic to a hydrodynamic regime can be readily tuned to occur \emph{through} the fluctuation-dominated ballistic regime, realizing a QCP-mediated nonequilibrium transition.

\section{Transport Setup}

Semiclassical transport is described by the Boltzmann equation that governs the evolution of a quasiparticle distribution $f(\mathbf{x}, \mathbf{p}, t)$ in the four-dimensional phase space of spatial $\mathbf{x} \equiv (x, y)$ and momentum $\mathbf{p} \equiv (p\sub{x}, p\sub{y})$ coordinates,
\begin{align}
\frac{\partial f}{\partial t} +  v\sub{F}\hat{\mathbf{p}}\cdot\frac{\partial f}{\partial \mathbf{x}} & = - \frac{f - f\super{mr}_{0} }{\tau\sub{mr}} -  \frac{f - f\super{mc}_0 }{\tau\sub{mc}}\label{eqn:fe_evol}
\end{align}
The terms on the right are MR and MC collision operators which relax $f$ to the local stationary $f\super{mr}_0(\mu\sub{mr})$ and shifted $f\super{mc}_0(\mu\sub{mc} + \mathbf{p}\cdot\mathbf{v}\sub{d})$ Fermi-Dirac distributions respectively, and are parametrized in a relaxation time approximation by the time scales $\tau\sub{mr}$ and $\tau\sub{mc}$. The local chemical potentials $\mu\sub{mr}(\mathbf{x})$ and $\mu\sub{mc}(\mathbf{x})$ enforce charge conservation for MR and MC scattering, and the local drift velocity $\mathbf{v}\sub{d}(\mathbf{x})$ enforces momentum conservation in MC scattering. The Ohmic regime occurs in the limit $\tau\sub{mr} \ll L/v\sub{F} \ll \tau\sub{mc}$ whereas the hydrodynamic regime requires $\tau\sub{mc} \ll L/v\sub{F} \ll \tau\sub{mr}$, where $L$ is a device scale. Finally, the ballistic regime is realized when both $\tau\sub{mr}, \tau\sub{mc} \gtrsim L/v\sub{F}$.

We solve (\ref{eqn:fe_evol}) at $T=0$ using the {\tt bolt} package (see methods). For concreteness, we consider the Fermi liquid in graphene ($v\sub{F}$ = 1 $\mu$m/ps). We take a rectangular geometry with dimensions $L\sub{x} \times L\sub{y} = L \times (\Gamma L + L\sub{c})$, where $\Gamma$ is the aspect ratio and $L\sub{c}$ is the width of the source/drain contacts through which transport is set up (fig.~\ref{fig:conductivity}a). The contacts inject/extract a small current ($\simeq 0.1$ $\mu$A) using a shifted Fermi-Dirac, and are perfectly Ohmic so as to exclude contact resistance \cite{imry-1998}. The device boundaries specularly reflect incident quasiparticles, needed to preserve the correlations that develop as quasiparticles traverse the device. Smooth boundaries have been seen in several ballistic transport experiments e.g., \cite{lee-2016, houten-1989, taychatanapat-2013}. We fix $L = 1$ $\mu$m and $L\sub{c} = 0.25$ $\mu$m. All devices are then parametrized by the aspect ratio $\Gamma$. A wire corresponds to $\Gamma = 0$. We label geometries with $0 < \Gamma \ll 1$ as ``caps'' and those with $\Gamma \gtrsim 1$ as ``wings''.

\section{Emergent dissipation} \label{sec:emergent_dissipation}

We first consider a wire ($\Gamma = 0$). In the absence of all bulk scattering $(\tau\sub{mr}, \tau\sub{mc}=\infty)$, the solution to (\ref{eqn:fe_evol}) is simply a spatially uniform shifted Fermi-Dirac, and thus zero resistance. We now slightly change the geometry by considering a cap ($\Gamma = 0.05$). Fig.~\ref{fig:conductivity}e shows the conductivity (computed using voltage measured across the source/drain contacts) as $\tau\sub{mr}$ is increased. For $\tau\sub{mr} \ll L/v\sub{F}$, the conductivity increases linearly in accordance with the Drude formula. However, the linear increase slows down for $\tau\sub{mr} \gtrsim L/v\sub{F}$. In the extreme limit of $\tau\sub{mr} = \infty$, the conductivity plateaus to a \emph{finite} value (dotted line in fig.~\ref{fig:conductivity}e). There is therefore a \emph{residual} resistance in the cap even for $\tau\sub{mr} = \infty$, as opposed to none in the wire.

To see what sets this residual resistance, we calculate the linear response for a circular Fermi surface in the $\tau\sub{mr} = \infty$ limit. We focus on the key feature of the cap geometry that differentiates it from a wire: the presence of a shear. Accordingly, we consider the $\sigma\sub{xx}$ component of the conductivity tensor for a spatial fluctuation with wavenumber $\mathbf{q} = (0, q\sub{y})$. The angular dispersion of velocities for quasiparticles on the Fermi surface gives rise to a pole in the integral for frequencies $\omega < q\sub{y}v\sub{F}$. In the DC limit (see methods),
\begin{align} 
\sigma(q\sub{y}) =  \frac{Ne^2}{h}\left(\frac{k\sub{F}}{q\sub{y}}\right) \label{eqn:sigma_xx}
\end{align}
Here $k\sub{F}$ is the Fermi wavevector and $N$ is the spin/valley degeneracy. We thus have a \emph{finite} conductivity even in the absence of all bulk scattering. An emergent dissipation occurs in device geometries that allow for shear ($q\sub{y}>0$). A wire only allows $q\sub{y} = 0$ and hence a zero resistance.

The ballistic conductivity (\ref{eqn:sigma_xx}) is set by the product $k\sub{F}\cdot l\sub{eff}$, where $l\sub{eff}$ is an effective scattering length determined only by the device (including contact) geometry. We now examine the dependence of the conductivity on the aspect ratio $\Gamma$ of our device geometry, first in the absence of disorder (fig.~\ref{fig:vortices}a). For $\Gamma \ll 1$ (caps), the resistance \emph{increases} $\propto \Gamma$ whereas it decreases in an Ohmic regime. Beyond $\Gamma \gtrsim 1$ (wings), the resistance saturates to a maximum and becomes \emph{independent} of the geometry. The underlying mechanism is visually striking (see section \ref{sec:vortices}). The saturated minimum conductivity equals that of a wire with length $L$ and a bulk scattering length $l\sub{eff} = 1.63 \pm 0.02$ $L$. The error is the difference between the equivalent scattering length for $\Gamma = 1$ and $\Gamma = 10$. Therefore, the wings have a \emph{universal conductivity} independent of nearly all microscopic details (band structure, interactions) as well as device dimensions,
\begin{align}
\sigma\sub{wing} = \frac{Ne^2}{2h}(k\sub{F} l\sub{eff}) && l\sub{eff} = 1.63 \pm 0.02\;L \label{eqn:sigma_xx_universal}
\end{align}
The relevant microscopic details for the conductivity (\ref{eqn:sigma_xx_universal}) are the shape of the Fermi surface (restricted here to circular) and the Fermi wavevector. With a finite disorder $\tau\sub{mr}\gtrsim L/v\sub{F}$, the resistance scales with device height in qualitatively the same way as the zero disorder case. It saturates for $\Gamma \gtrsim 1$ and is offset compared to the corresponding value at zero disorder (fig.~\ref{fig:vortices}(a)).

\section{Spatial Fluctuations: Vortices} \label{sec:vortices}

The ballistic regime is usually viewed in terms of individual quasiparticle trajectories. A surprising aspect is that the currents organize themselves into vortices; a \emph{collective} effect. The current has a finite \emph{vorticity} $\omega \equiv \hat{z}\cdot\left(\nabla \times \mathbf{v}\sub{d}\right)$, where $\hat{z}$ is the direction perpendicular to the two-dimensional Fermi liquid. An applied field of the form $E\sub{x}(q\sub{y})$, produced by device geometries with $\Gamma > 0$, shears the Fermi fluid. The shear produces a vorticity $\omega(q\sub{y}) = \Omega(q\sub{y})E\sub{x}(q\sub{y})$, where $\Omega(q\sub{y}) \equiv iq\sub{y}\sigma\sub{xx}(q\sub{y})/en $ is the vortical response to the shear. The \emph{form} of the conductivity determines the shear response. The scale-dependent ballistic conductivity (\ref{eqn:sigma_xx}) gives 
\begin{align}
\Omega(q\sub{y}) & = i\frac{2e}{\hbar}\frac{1}{k\sub{F}} \label{eqn:vorticity_ballistic}
\end{align}
The ballistic regime has a finite scale-invariant vortical response to shear, and is determined only by the Fermi wavenumber. In contrast, the Ohmic regime has $\Omega(q\sub{y}) = (ie/m)q\sub{y}\tau\sub{mr}$, which vanishes for $q\sub{y}\rightarrow 0$. While a finite vortical response is necessary to produce vortices, it is not sufficient. Vortices also require a conducive device and contact geometry. The geometry we present is one such example (counterexamples in supplementary fig.~\ref{fig:asymm_contact}). Given sufficient room ($\Gamma \gtrsim 0.2$), the current responds to the external field by twisting into vortices (inset in fig.~\ref{fig:vortices}a).

Vortices are ``bound states'' of the current with zero net transport. The emergent dissipation in our geometry and its saturation to the universal value (\ref{eqn:sigma_xx_universal}) can be directly visualized as an excitation of these current bound states. The caps ($\Gamma < 0.2$) have no vortices and a \emph{geometry dependent} dissipation $\propto \Gamma$. At $\Gamma = 0.25$, vortices appear at the top corners and the resistance ceases to increase. As $\Gamma$ is further increased, the corner vortices grow and coalesce to give a single dominant vortex above the source-drain axis. This dominant vortex always persists for $\Gamma > 1$, with ever smaller vortices emerging from the top edge (fig.~\ref{fig:vortices}b). These smaller vortices contribute negligibly to the overall resistance, and it becomes \emph{independent} of the geometry.

The scale-invariant ballistic response (\ref{eqn:vorticity_ballistic}) produces visually striking current flow patterns for $\Gamma \gg 1$, with vortices prevalent at all scales. Such copious vortex generation in two-dimensions is highly non-classical. For comparison, we show the flow structure in a strongly hydrodynamic regime ($\tau\sub{mc} = 0.01$ ps) where there is only one vortex at the device scale (fig.~\ref{fig:vortices}c). A detailed comparison between the ballistic and hydrodynamic vortical response is presented in section~\ref{sec:hydro_ballistic}, which we summarize here. The hydrodynamic response peaks at low spatial wavenumbers (and hence only a single vortex is seen), disappears as $T \rightarrow 0$,  and is highly susceptible to disorder. The ballistic response persists even at $T = 0$ and is robust to disorder, with vortices appearing as soon as $\tau\sub{mr} \gtrsim L/(2v\sub{F})$ (fig.~\ref{fig:conductivity}a-d).

\section{Regime Transitions and Fermi Surface fluctuations} \label{sec:fermi_surface}

We now study the transitions between all three transport regimes using a mode decomposition of the Fermi surface. The Fermi surface at equilibrium, denoted by $f_0$, is a perfect circle. The circle deforms/shifts in the presence of an applied field. The perturbation $\delta f$ of the resulting nonequilibrium distribution $f = f_0 + \delta f$ can be decomposed into angular fluctuations,
\begin{align}
\delta f(\mathbf{x}, \theta) & = \sum_{n = 0}^{\infty} \delta f_n(\mathbf{x}) \cos(n\theta + \phi_n) \\
& \equiv A(\mathbf{x}) + B(\mathbf{x}) \cos(\theta + \phi_1) + \Delta f(\mathbf{x}, \theta) \label{eqn:fourier_series}
\end{align}
where $\delta f_n$ is the (real) amplitude of a fluctuation with discrete wavenumber $n$ and $\phi_n$ is the associated phase. We pull out the $n=0, 1$ modes in (\ref{eqn:fourier_series}). The amplitude $A = \mu - \mu\sub{F}$ represents a change in the radius of the circle, with $A > 0$ ($A < 0$) corresponding to electron (hole) overdensities. $B = p\sub{F}|\mathbf{v}\sub{d}|$ is the magnitude of a shift in the origin, in the direction $\phi_1$. The rest of the terms in (\ref{eqn:fourier_series}), denoted by $\Delta f$, are small scale angular fluctuations (visualized in the supplementary fig.~\ref{fig:modes_schematic}).

The distributions for each of the regimes are shown for a wing with $\Gamma = 1$ in fig.~\ref{fig:dist_func}, at a spatial location above the source-drain axis. The Ohmic and hydrodynamic regimes have local equilibrium distributions that satisfy $|A| \gg B \gg |\Delta f|$ and $B \gg |A| \gg |\Delta f|$ respectively. The fluctuations in both regimes are negligible, allowing for a mean-field treatment using only the variables $\{A, B, \phi_1\}$. In contrast, the ballistic distribution is \emph{dominated} by fluctuations, $|\Delta f|\gg |A|, B$. It has a proliferation of holes ($\delta f < 0$) interspersed with electrons ($\delta f > 0$) at \emph{all} angular scales.

There are two transitions possible as $\tau\sub{mr} \ll L/v\sub{F}$ is increased to $\tau\sub{mr} \gtrsim L/v\sub{F}$. Depending on the relative magnitude between $\tau\sub{mc}$ and $\tau\sub{mr}$, we get a ``second-order'' fluctuation mediated Ohmic-Ballistic-Hydrodynamic or a ``first-order'' Ohmic-Hydrodynamic transition. We illustrate both for a wing with $\Gamma = 1$. We start in an Ohmic regime and reach the hydrodynamic regime \emph{through} the ballistic regime. We then take the first-order route and directly go back to the Ohmic regime.

\underline{Ohmic$\rightarrow$Ballistic$\rightarrow$Hydrodynamic}: We begin in fig.~\ref{fig:dist_func}f with a fixed $\tau\sub{mc} = 100$ ps, and increase $\tau\sub{mr}$ from $0.01$ ps to $100$ ps. For small $\tau\sub{mr}$, MR scattering suppresses all modes apart from the local equilibrium mode $A$. As $\tau\sub{mr}$ increases, the suppression becomes ineffective and the fluctuations $|\Delta f|$ grow. They quickly overcome $A$, and saturate beyond $\tau\sub{mr} \gtrsim L/v\sub{F} = 1$ ps to give the ballistic distribution in fig.~\ref{fig:dist_func}b. Now keeping $\tau\sub{mr} = 100$ ps, we decrease $\tau\sub{mc}$ from $100$ ps to $0.01$ ps in fig.~\ref{fig:dist_func}g. The MC interactions eventually thermalize all the fluctuations and we get the local equilibrium $B$ of the hydrodynamic regime (fig.~\ref{fig:dist_func}c).

\underline{Hydrodynamic$\rightarrow$Ohmic}: We now fix $\tau\sub{mc} = 0.01$ ps and decrease $\tau\sub{mr}$ from $100$ ps to $0.01$ ps in fig.~\ref{fig:dist_func}h. The constant presence of strong bulk MC scattering inhibits fluctuations. The hydrodynamic regime then directly transitions into an Ohmic regime when MR scattering is sufficiently strong. The Fermi surface is a local equilibrium throughout, changing abruptly from $B$ to $A$.

\section{Experimental Probes and Critical Scaling} \label{sec:critical_scaling}

The relativistic CFT underlying the ballistic regime is invariant under a scaling transformation with $z = 1$. We show how this critical scaling manifests in transport. The scale-dependent ballistic conductivity produces a \emph{negative} nonlocal resistance in DC transport, and a \emph{positive} nonlocal phase in AC transport\cite{chandra-2019}\footnote{\emph{Any} scale-dependent conductivity has these signatures, leading to a degeneracy between the ballistic and the hydrodynamic regime. However, these regimes can be distinguished by checking for the dominance of bulk interactions using spatiotemporal voltage-voltage correlations in AC transport \cite{chandra-2019}.}; a nonlocal voltage leads the source current. These probes allow us to identify the quantum critical ballistic regime in the $(\tau\sub{mr}, f\sub{source})$ parameter space, where $f\sub{source}$ is the source frequency. We assume temperatures are low enough that MC interactions are negligible ($l\sub{mc} \sim T^{-2} > L$).

\begin{figure*}[!htbp]
\begin{center}
\includegraphics[width=160mm]{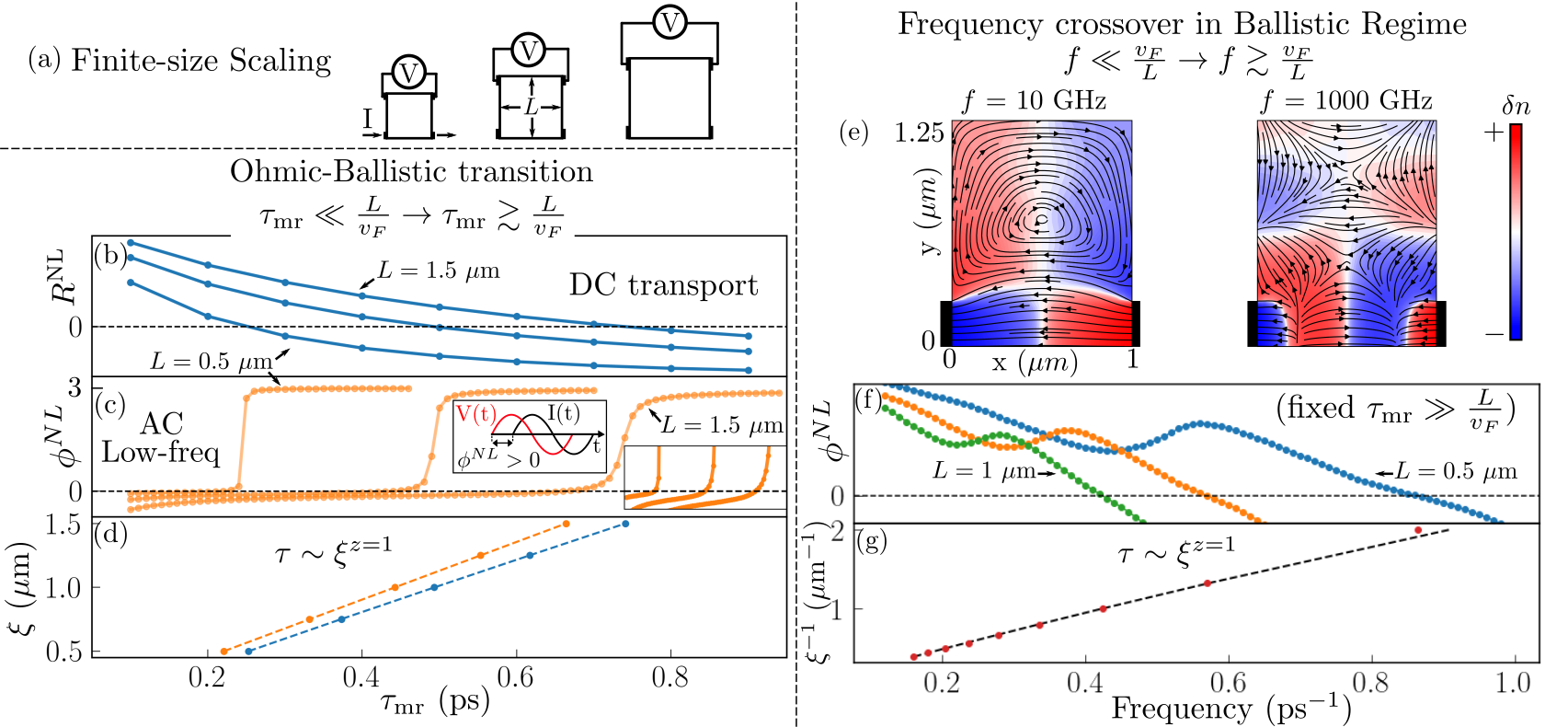}
\end{center}
\caption{\emph{Experimental probes and critical scaling}: (a) Finite-size scaling setup to demonstrate critical scaling of the ballistic regime, and extract $z$. The geometry is a wing with $\Gamma =1$. (b) Nonlocal resistance $R^{NL}$, computed using voltage measured at the top edge of the device (as shown in (a)), versus $\tau\sub{mr}$ for various device scales $L$. An Ohmic regime has $R^{NL}>0$, whereas the ballistic regime has $R^{NL}<0$. (c) The phase $\phi^{NL}$ between the nonlocal voltage measured at the top edge, and the current source in low-frequency AC transport ($f\ll v\sub{F}/L$). An example ballistic time series, where $\phi^{NL}>0$, is shown in inset. An Ohmic regime has $\phi^{NL}<0$; the measured voltage lags the source. (d) The spatiotemporal correlation scales, which fit to $\tau = C \xi^z$ where $C \simeq 1/(2v\sub{F})$ and $z=1$. The correlation time $\tau$ is measured in DC (blue) using $\tau = \tau\sub{mr}(R^{NL}=0)$, and in AC (orange) using $\tau = \tau\sub{mr}(\phi^{NL}=0)$. The corresponding correlation length $\xi$ is the device scale $L$, since the probes are at the device edge (shown in (a)). (d) Current streamlines and density fluctuations in low-frequency (left) and high-frequency (right) transport, at the same temporal location ($t = 0.75 T\sub{0}$, $T\sub{0}\equiv$ time period) of the AC cycle. (e) $\phi^{NL}$ versus frequency for various device scales. The ballistic regime (with $\tau\sub{mr} = 5$ ps $\gg L/v\sub{F}$) crosses over from a low-frequency collective regime ($\phi^{NL}>0$) to a high-frequency collisionless regime ($\phi^{NL}<0$). The low- and high-frequency dynamics, and the flow profile across the frequency crossover is shown in the movie \footnote{\url{https://vimeo.com/366725650}}. (f) Spatiotemporal scales obtained from the frequency crossover in (e), using the same AC technique as in (c). Here $\tau = 1/f(\phi^{NL}=0)$. The scales fit to $\tau = C_1 \xi^z $, with $C_1\simeq 2.2/v\sub{F}$ and $z = 1$.}\label{fig:critical_scaling}
\end{figure*}

\underline{Ohmic-Ballistic transition}: We consider the Ohmic-Ballistic transition in a wing with $\Gamma = 1$. The transition, produced by tuning $\tau\sub{mr}$, is detected in DC transport using a nonlocal resistance $R\super{NL}$ measured at the edge of the device. $R\super{NL}$ crosses over from positive in the Ohmic to negative in the ballistic regime (fig.~\ref{fig:critical_scaling}b). Similarly, a nonlocal phase $\phi\super{NL}$ in low-frequency AC transport ($f\sub{drive}  = 10$ GHz $\ll v\sub{F}/L$) changes from negative in the Ohmic to positive in the ballistic regime (fig.~\ref{fig:critical_scaling}c). Precisely when $R\super{NL}=0$ (or $\phi\super{NL} = 0$), the correlation length $\xi = L$ and the correlation time $\tau = \tau\sub{mr}$. By repeating this procedure for devices of varying sizes, we get the corresponding spatiotemporal correlation scales which obey $\tau = C\xi^z$ with $C \simeq 1/(2v\sub{F}), z=1$ (fig.~\ref{fig:critical_scaling}d).

\underline{Ballistic frequency crossover}: For this setup, $\tau\sub{mr}$ is fixed ($\gg L/v\sub{F}$) such that the device is always in the ballistic regime. However, the correlations induced by ballistic quasiparticles require a finite time $\sim L/v\sub{F}$ to traverse the device. As $f\sub{source}$ is increased, there is a critical frequency $f\sub{c}$ beyond which the source changes faster than the time it takes for device to be correlated. The collective features of the ballistic regime such as vortices disappear (fig.~\ref{fig:critical_scaling}e), and the device transitions into the usual high-frequency ``collisionless'' regime ($f\sub{source} \gg 1/\tau\sub{mr}, v\sub{F}/L$)\cite{AshMer}. The latter has a scale-independent conductivity and thus a negative nonlocal phase. Precisely at $f\sub{c}$, we have $\phi\super{NL}=0$. The correlation time is therefore $\tau = f\sub{c}^{-1}$ and the correlation length $\xi = L$, which fit to $\tau = C_1\xi$ where $C_1 \simeq 2.2/v\sub{F}, z = 1$ (fig.~\ref{fig:critical_scaling}g).

\section{Discussion}

The Fermi gas has long been considered a trivial QCP\cite{sachdev-1995}; a viewpoint vindicated by ballistic wire transport. However, its true character emerges in a wing (fig.~\ref{fig:vortices}b), which probes the shear response of a many-body system. Ballistic transport now exhibits a universal \emph{intrinsic} dissipation (section \ref{sec:emergent_dissipation}), distinctive fluctuations (sections \ref{sec:vortices} and \ref{sec:fermi_surface}) and obeys critical scaling (section \ref{sec:critical_scaling}); all characteristics of a QCP. A quantum critical viewpoint of the ballistic regime provides new insights.

It is well-known that just boundary scattering can give rise to a dissipation, given by the Landauer-B\"{u}ttiker formula (see supplementary section~\ref{sec:particle_methods}). By invoking criticality, we are able to endow a sense of universality to this dissipation and \emph{see} the emergence of this universality in terms of well-defined \emph{bulk} fluctuations; vortices. The current and density fluctuations can be directly imaged using scanning probe techniques\cite{ella-2019}. Crucially, the critical scaling $\tau\sub{bulk} \sim L^z$ controls the requirements on MR scattering; if the underlying CFT had $z = 2$ for example, the regime would be much harder to access.

Vortices in a wing with $\Gamma \gg 1$ directly count conservation laws. A Fermi gas has an infinite number of conserved momenta which we \emph{see} as vortices at multiple scales (fig.~\ref{fig:vortices}b). In contrast, a hydrodynamic regime only conserves a single coarse-grained momentum and we indeed see just a single vortex at the device scale (fig.~\ref{fig:vortices}c). This argument will prove useful in deducing the flow structure of transport in \emph{strongly-interacting} QCPs, where quasiparticles are not well-defined, such as in graphene at charge neutrality \cite{sheehy-2007, son-2007}. A strongly-interacting hydrodynamic regime emerges at such QCPs \cite{andreev-2011, crossno-2016, gallagher-2019, lucas-2016, muller-2009}. The regime has a single macroscopically conserved momentum, and we expect to see a single vortex as in fig.~\ref{fig:vortices}c.

We thus argue that while strongly-interacting QCPs are fluctuation dominated in the single-particle Green's function, they have a qualitatively simple transport profile. On the contrary, an interaction-free QCP with well-defined quasiparticles has transport involving fluctuations in \emph{each} dimension of the quasiparticle distribution $f(t, \mathbf{x}, \mathbf{p})$. The temporal fluctuations in DC transport merit an independent discussion and will be presented in a forthcoming paper.

\newpage

\begin{figure*}[!htbp]
\begin{center}
\includegraphics[width=160mm]{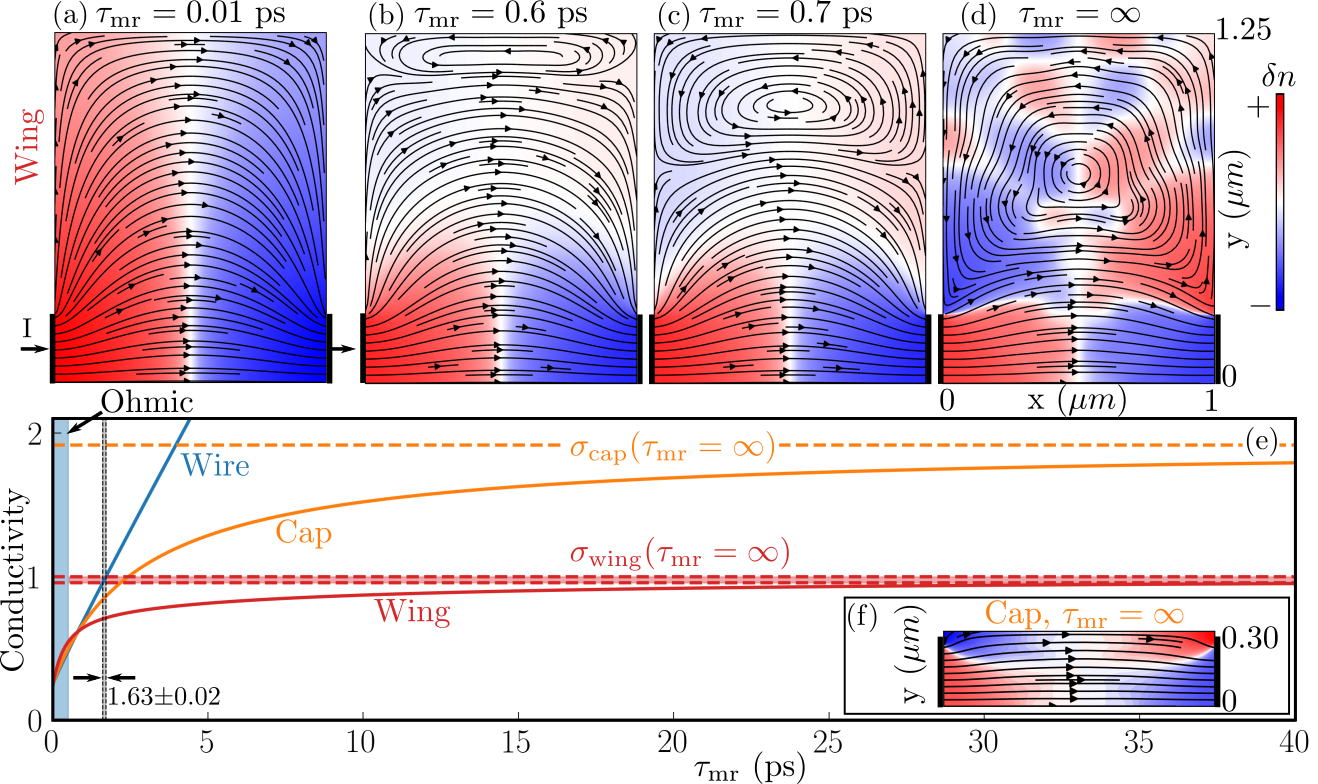}
\end{center}
\caption{\emph{Emergent dissipation}: We consider devices with dimensions $L\sub{x}\times L\sub{y} = L \times (L\sub{c} + \Gamma L)$, where $L = 1$ $\mu$m and the contact width $L\sub{c} = 0.25$ $\mu$m. The aspect ratio $\Gamma$ specifies the device: a \emph{wire} has $\Gamma = 0$, \emph{caps} have $0 < \Gamma \ll 1$ (shown in inset (f) for $\Gamma = 0.05$) and \emph{wings} have $\Gamma \gtrsim 1$ (shown in (a-d) for $\Gamma = 1$). (a-d) Current streamlines and density fluctuations ($\delta n$) in a wing with $\Gamma = 1$. For a bulk momentum-relaxing scattering timescale $\tau\sub{mr}=0.01$ ps $\ll L/v\sub{F} = 1$ ps, the flow is Ohmic. The ballistic regime sets in as soon as $\tau\sub{mr} \gtrsim L/(2v\sub{F}) = 0.5$ ps: vortices appear from the top and signal the onset of an emergent \emph{intrinsic} dissipation. At $\tau\sub{mr} = \infty$, only this intrinsic dissipation remains. (e) Conductivity vs $\tau\sub{mr}$ for a wire, a cap ($\Gamma = 0.05$) and a wing ($\Gamma = 1$). The conductivity $\sigma$ for a wire increases without bound as $\tau\sub{mr} \rightarrow \infty$, whereas the cap and the wing saturate to finite limits (shown in dotted lines). The dissipation in wings is \emph{independent} of geometry: the two dotted red lines correspond to $\sigma\sub{wings}(\tau\sub{mr} = \infty)$ for wings with $\Gamma = 1$ (upper line) and $\Gamma = 10$ (lower line). The equivalent conductivity of a wire occurs at $\tau\sub{mr}$ = (1.63 $\pm$ 0.02) $L/v\sub{F}$. Note the small extent of the Ohmic regime (shaded in blue) in the $\tau\sub{mr}$ parameter space, where $\partial\sigma/\partial \Gamma > 0$ and is simply due to addition of parallel current paths.}\label{fig:conductivity}
\end{figure*}

\newpage
\begin{figure*}[!htbp]
\begin{center}
\includegraphics[width=160mm]{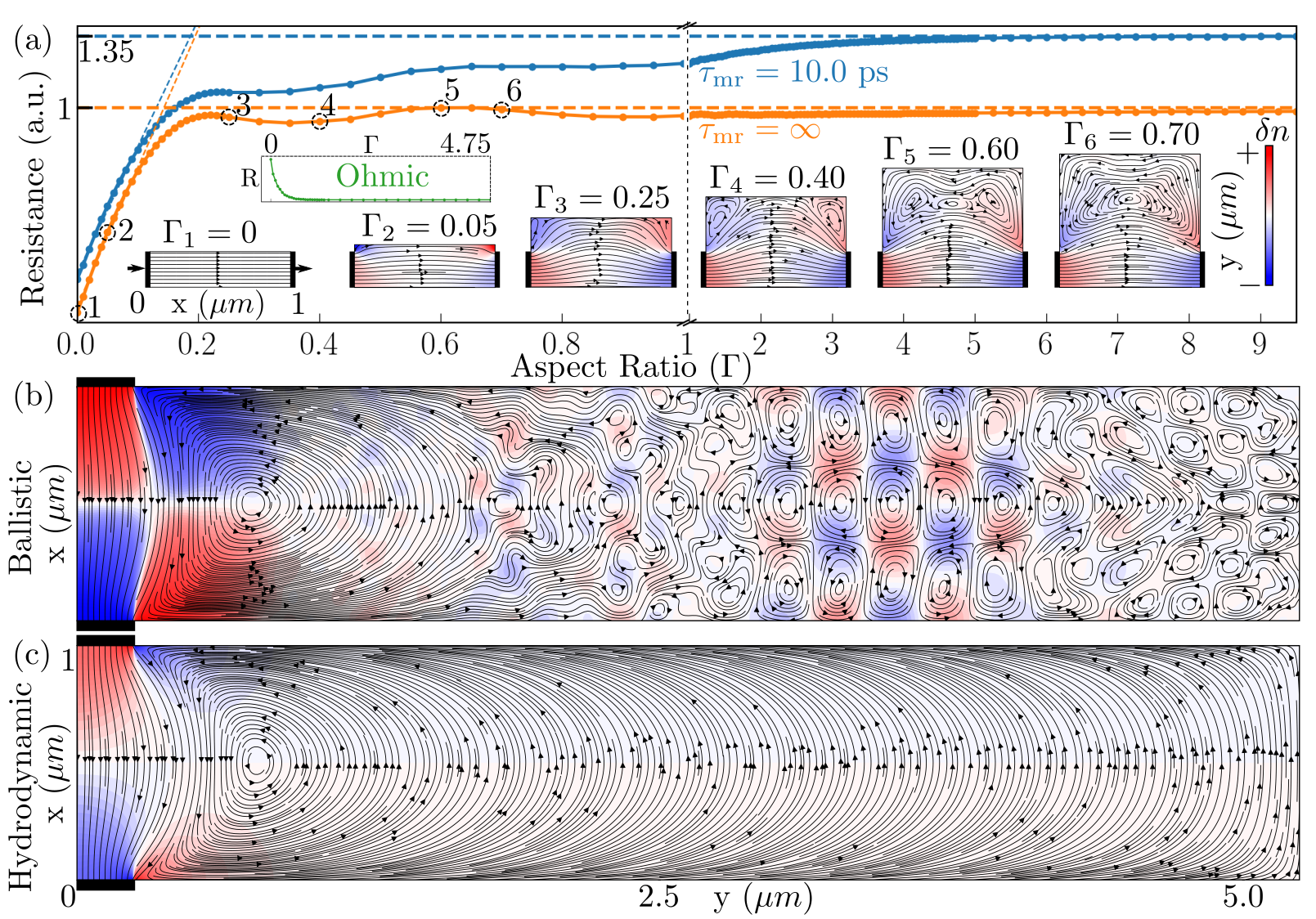}
\end{center}
\caption{\emph{Spatial fluctuations}: (a) Resistance $R$ vs aspect ratio $\Gamma$. First consider no bulk momentum-relaxing scattering ($\tau\sub{mr}=\infty$), shown in orange. The resistance starts from zero for a wire ($\Gamma = 0$) and \emph{increases} linearly (fit shown using dotted line) in caps ($\Gamma \lesssim 0.2$). The ballistic caps thus have $\partial R/\partial \Gamma > 0$ whereas Ohmic caps have $\partial R/\partial \Gamma < 0$ (inset). As $\Gamma$ is further increased, vortices appear and the resistance \emph{saturates} in wings ($\Gamma \gtrsim 1$) to the universal value (\ref{eqn:sigma_xx_universal}). The transient vortical structures (shown in insets) leading up to the single dominant vortex of a wing (see fig.~\ref{fig:conductivity}d) result in small resistance undulations. The results are robust to a finite $\tau\sub{mr} \gtrsim L/v\sub{F}$ (shown in blue for 10 ps), with extrinsic dissipation contributing to the saturated resistance . (b) As $\Gamma \rightarrow \infty$, smaller vortices break out of the dominant vortex resulting in a striking spatial profile with fluctuations at all scales (shown here for $\Gamma = 5$). Note the tight correlation between the current vortices and the density fluctuations. The emergence of these spatial fluctuations is shown in the movie \footnote{\url{https://vimeo.com/365020115}}. (c) The smooth flow profile of a hydrodynamic regime (shown for the same geometry as (b)), where momentum-conserving scattering dominates (here $\tau\sub{mc}=0.01$ ps, $\tau\sub{mr}=100$ ps). Note the contrast to the ballistic flow (b).}\label{fig:vortices}
\end{figure*}

\newpage
\begin{figure*}[!htbp]
\begin{center}
\includegraphics[width=160mm]{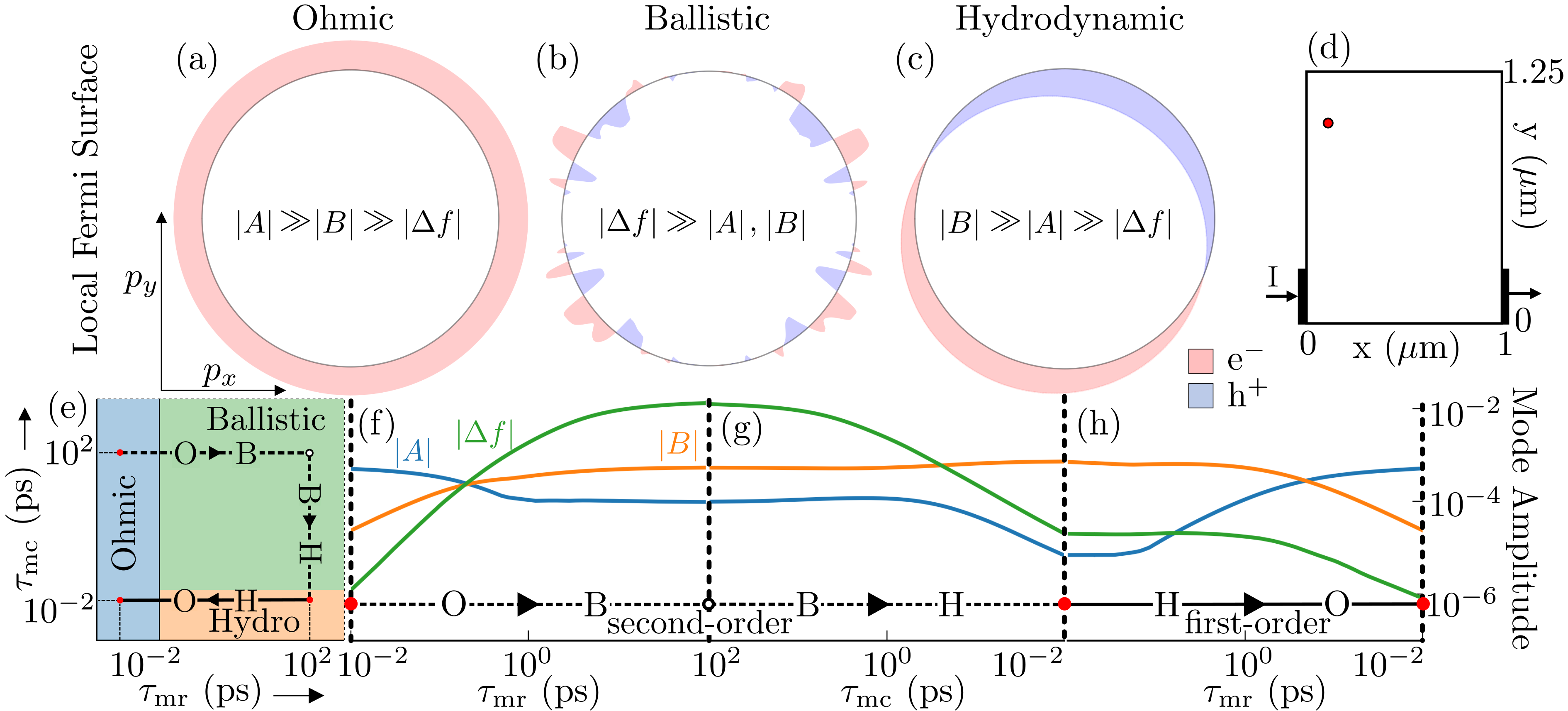}
\end{center}
\caption{\emph{Regime transitions and Fermi surface fluctuations}: (a-c) The local Fermi surface for different regimes in a wing with $\Gamma =1$, shown for a typical spatial location (marked by a red dot in (d)). The black circle denotes the background equilibrium, with an overdensity of electrons (holes) shaded in red (blue). The mode amplitude $|A|$ corresponds to a change in the radius of the circle, $|B|$ indicates a shift in the center, and $\Delta f$ is the sum of small scale angular fluctuations. The Ohmic distribution is dominated by $|A|$, with an imperceptible shift $|B|>0$. In contrast, the hydrodynamic distribution has a dominant shift. The ballistic distribution is fluctuation dominated, with scale-invariant electron-hole pockets (see also fig.~\ref{fig:fermi_surface}). (e) Schematic showing the path we take in the $(\tau\sub{mr}, \tau\sub{mc})$ parameter space to traverse the different regimes. (f-h) The device averaged mode amplitudes along each path in (e). The vertical dotted lines mark the end points of each path, which is where the distributions in (a-c) are shown. The regime transitions are shown in the movie \footnote{\url{https://vimeo.com/364982637}}. The Ohmic-Ballistic-Hydrodynamic transition (O$\rightarrow$B and B$\rightarrow$H in (e)) occurs continuously, whereas the Hydrodynamic-Ohmic transition is abrupt (H$\rightarrow$O in (e))}\label{fig:dist_func}
\end{figure*}

\newpage

\newpage

\section{Methods}

\subsection{Numerical scheme for the Boltzmann equation}

We solve (\ref{eqn:fe_evol}) using a high-resolution finite volume scheme implemented in {\tt bolt}\cite{chandra-bolt}, a fast GPU accelerated solver for kinetic theories. The code timesteps the quasiparticle distribution in a discrete $(\mathbf{x}, \mathbf{p})$ domain, with error $\mathcal{O}(\Delta x^2, \Delta p^2, \Delta t^2)$, where $\Delta x$, $\Delta p$ and $\Delta t$ are the spatial, momentum and temporal grid spacings respectively. We work in the $T = 0$ limit where the Fermi surface is confined to a one-dimensional line within the two-dimensional momentum space, resulting in a significant speedup. We typically use 72 grid zones per micron along each spatial axis and 8192 zones on the Fermi surface, for a total of $\simeq 50\Gamma \times 10^6$ unknowns ($\Gamma \equiv$ aspect ratio). All simulations are initialized with a thermal distribution (background density $\simeq 10^{12}$ cm$^{-2}$) and evolved to a steady state ($\gtrsim$ 100 ps) with a timestep $6.25 \times 10^{-3}$ ps. Note that a chemical potential gradient, due to current injection/extraction at the source/drain contacts, appears as a long-range force in (\ref{eqn:fe_evol}) at linear order \cite{chandra-2019}.

The two-dimensional devices are assumed to be in a field-effect transistor (FET) configuration wherein the spatial density gradients and the in-plane voltage fluctuations are related by a geometric capacitance. This relationship in FETs, known as the local capacitance approximation (LCA)\cite{tomadin-2013}, is used in section (\ref{sec:critical_scaling}) to obtain the voltage from a spatial density profile, and thus the current-voltage relation. The scheme is described in \cite{chandra-2019}.

The high-resolution second-order algorithm is essential to resolving the critical fluctuations shown in the main text. Ballistic transport is typically simulated using Monte-Carlo particle methods which have a much higher noise floor. We compare the performance of our deterministic algorithm against particle methods in the supplementary (fig.~\ref{fig:particle_trajectories}).

\subsection{Conductivity}
The conductivity in two-dimensions for a frequency $\omega$, a spatial wavenumber $\mathbf{q}$ and a momentum relaxing time scale $\tau\sub{mr}$ is given by,
\begin{align} \label{eqn:cond_2D_defn}
\sigma_{ij}(\mathbf{q}, \omega) & = \frac{Ne^2}{m}\int \frac{d^2p}{4 \pi^2 \hbar^2}\left(-\frac{\partial f}{\partial p_j} \right) \frac{p_i(\mathbf{k})}{(1/\tau\sub{mr}) + i\left(-\omega + \mathbf{q}\cdot \mathbf{v}(\mathbf{p})\right)}
\end{align}
where $m \equiv p\sub{F}/v\sub{F}$ is an effective mass, $p\sub{F} = \hbar k\sub{F}$ is the Fermi momentum, $k\sub{F}$ is the Fermi wavevector, $v\sub{F}$ is the Fermi velocity and $N$ is the number of spins and valleys. We consider the $(i, j) = (x, x)$ component with $\mathbf{q} = (0, q\sub{y})$. At $T = 0$, $-\partial f/\partial p_x = \delta(p - p\sub{F})\hat{p}\sub{x}$ and $\mathbf{v}(\mathbf{p}) = v\sub{F}\hat{p}$, where $\mathbf{p} = p\sub{F}\left(\cos(\theta), \sin(\theta) \right)$. Finally, setting $\tau\sub{mr} = \infty$, we have,
\begin{align}
\sigma_{xx}(q\sub{y}, \omega) & = \frac{Ne^2}{4 \pi^2 \hbar} \left(\frac{k\sub{F}}{q\sub{y}}\right) \mathcal{I}\left(\frac{\omega}{q\sub{y}v\sub{F}}\right) \label{eqn:conductivity_in_terms_of_I}
\end{align}
where,
\begin{align}
\mathcal{I}(x) & = i\lim_{\epsilon \rightarrow 0^+}\int_0^{2\pi}d\theta \frac{\cos(\theta)^2}{x + i\epsilon - \sin(\theta)} \label{eqn:integral}
\end{align}
There are now two distinct frequency regimes: (1) for $\omega \gg q\sub{y}v\sub{F}$, the integrand in (\ref{eqn:integral}) has no poles. The resulting integral $\mathcal{I}$ is purely imaginary and we recover the usual collisionless regime where $\sigma\sub{xx}(q\sub{y}, \omega) \sim i/\omega$. (2) For $\omega < q\sub{y}v\sub{F}$, the presence of a pole in the integrand requires taking the limit of vanishing $\epsilon \sim 1/\tau\sub{mr}$  (or equivalently, a sufficiently large $\epsilon \equiv \mathrm{Im}(\omega)>0$ so that the Laplace transforms, $A(\omega) = \int_0^\infty A(t)e^{i \omega t} dt$, of all fields converge as $t \rightarrow \infty$). The integral in the DC limit is $\mathcal{I}(0) = 2\pi$, resulting in a purely real conductivity in (\ref{eqn:conductivity_in_terms_of_I}). The result can be cast in a Drude form, $\sigma\sub{xx} = ne^2\tau\sub{eff}/m$ (density $n = N k\sub{F}^2/(4\pi)$, mass $m = \hbar k\sub{F}/v\sub{F}$), with an effective scattering time scale $\tau\sub{eff} = 2/(v\sub{F}q\sub{y})$. The ballistic conductivity can also be obtained from a diagrammatic evaluation of the current-current correlation, without any reference to the Boltzmann equation\cite{glasser-1963}.

The conductivity is thus scale-dependent and has a nonlocal current-voltage relation for $\tau\sub{mr}, 1/\omega > L/v\sub{F}$, and becomes scale-independent (local current-voltage relation) when either $\tau\sub{mr} < L/v\sub{F}$ or $\omega > v\sub{F}/L$. We show critical scaling across both transitions in section (\ref{sec:critical_scaling}).

\renewcommand\thefigure{S.\arabic{figure}}
\renewcommand\thesection{S}
\setcounter{section}{0}
\section{Supplementary}
\setcounter{figure}{0}

\subsection{Landauer-B\"{u}ttiker formalism and Particle methods} \label{sec:particle_methods}

In mesoscopic devices operating in the semiclassical ballistic regime, dissipation is understood within the ``billiards ball'' interpretation of the Landauer-B\"{u}ttiker formalism \cite{beenakker-1989}. We show that this viewpoint is completely consistent with, and complementary to the phase-mixing picture of emergent dissipation. First consider the wire. All electrons injected through the source, moving along their classical trajectories, reach the drain. The transmission $t$ is therefore unity, and the resistance $\propto (1 - t)/t$ as per the Landauer-B\"{u}ttiker formula is zero.

\begin{figure*}[!htbp]
\begin{center}
\includegraphics[width=160mm]{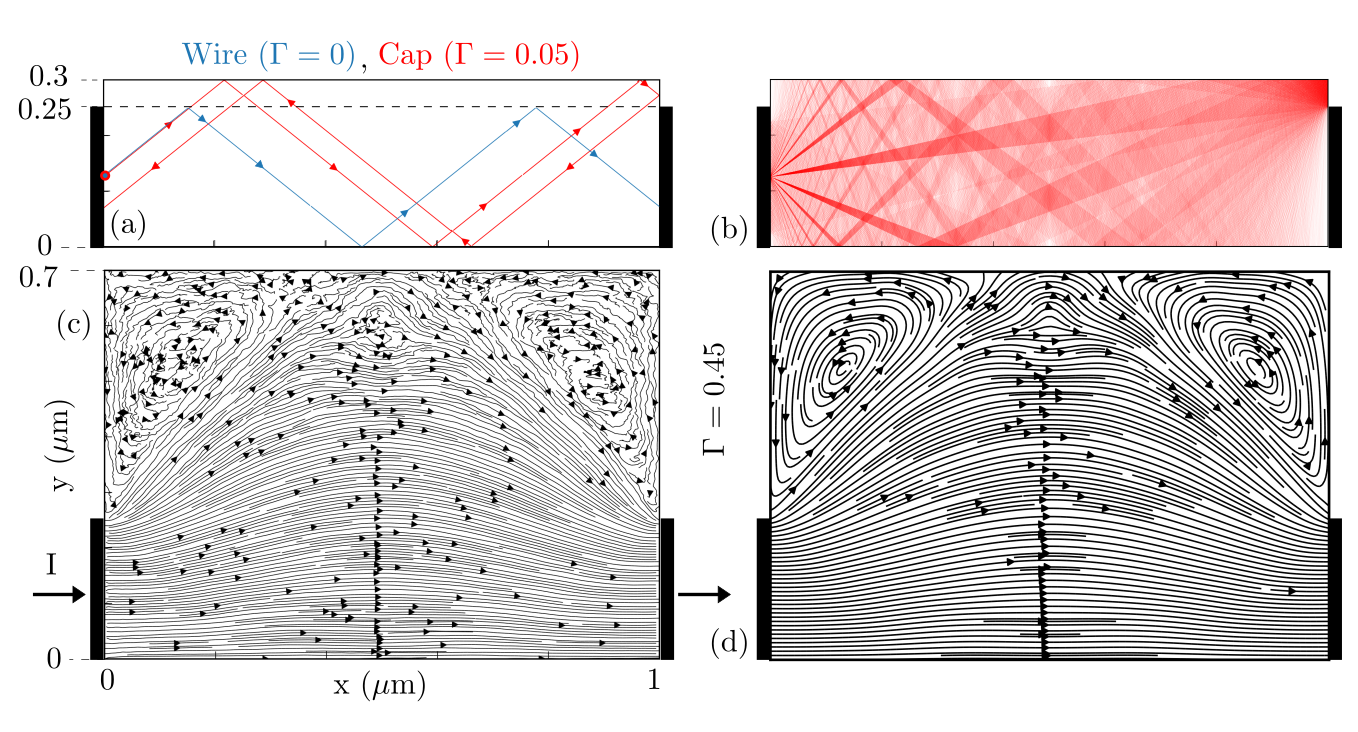}
\caption{a) Particle trajectories from a point on the source contact, marked by a red dot, for a specific injection angle. The blue line shows the trajectory in a wire (delimited by the dotted line), and the red line in a cap. The trajectories in a wire, for all injection angles, lead to the drain. The small gap between the top edge and the contact in the cap gives rise to closed trajectories for certain injection angles, as shown in the example here; the trajectory leads back into the source. (b) All closed trajectories originating from a point on the source contact (red dot in (a)). They produce a particle overdensity at the top right corner (same as red regions in fig.~\ref{fig:conductivity}f), and thus a negative nonlocal resistance. (c) Current streamlines obtained by computing the vector sum of all particle velocities. Note the presence of vortices. (d) Current streamlines obtained by solving the Boltzmann equation, using {\tt bolt}. Both the particle simulation, as well as the Boltzmann solution give rise to vortices, with the noise floor being much lower for {\tt bolt}.}\label{fig:particle_trajectories}
\end{center}
\end{figure*}

Now consider the cap. We have seen in the main text that the Boltzmann equation predicts a finite dissipation in the presence of a shear, even with \emph{zero} bulk scattering. This dissipation arises from the phase-mixing term $v\sub{F}\hat{\mathbf{p}}\cdot\partial f/\partial \mathbf{x}$ in the Boltzmann equation (\ref{eqn:fe_evol}). It can also be seen as arising from boundary scattering in the billiards picture. The region perpendicular to the contacts gives rise to a finite number of closed trajectories: some quasiparticles injected from the source contact follow trajectories that lead back into the source, instead of into the drain (fig.~\ref{fig:particle_trajectories}a). The transmission is now no longer unity and thus a finite resistance. These closed trajectories also give rise to a \emph{negative} nonlocal resistance (fig.~\ref{fig:particle_trajectories}b), which indeed agrees with the Boltzmann solution (fig.~\ref{fig:conductivity}f), and is a signature of a nonlocal current-voltage relation.

In the main text, we have shown the presence of vortices by directly solving the Boltzmann equation. We now show that a computation of currents using the billiards ball model, i.e.~single-particle trajectories reflecting off the device boundaries, also yields the same result. We consider a device with an aspect ratio $\Gamma = 0.45$. The source contact of width $L\sub{c} = 0.25$ $\mu$m is discretized uniformly using $N\sub{c}$ gridpoints. Each source gridpoint $i$ injects particles with velocity $v\sub{F}$, at discrete angles $\theta_j$ which sample the angular domain $[-\pi/2, \pi/2]$ at $N_\theta$ equispaced points. The particles then follow a classical trajectory $\mathbf{x}_{ij}(t)$ with velocity $\mathbf{v}_{ij}(t)$ till they exit the device at time $t = t^{ij}$, either via the source or the drain contacts. Note that each individual trajectory, indexed by $ij$, has a different exit time $t^{ij} = N_t^{ij}\Delta t$, where $\Delta t$ is the timestep and $N^{t}_{ij}$ is the number of discrete timesteps taken by the trajectory. The currents are obtained by summing over the velocities along each single-particle trajectory,
\begin{align}
\mathbf{J}(\mathbf{x}) = \sum\limits_{i=1}^{N\sub{c}}\sum_{j=1}^{N_\theta}\sum_{n=1}^{N_t^{ij}} P(\theta_j) \mathbf{v}_{ij}(n\Delta t) \delta(\mathbf{x} - \mathbf{x}_{ij}(n \Delta t))
\end{align}
The factor $P(\theta) = 0.5 \cos(\theta)$ is the angular distribution of the injected particles\cite{beenakker-1989}. It corresponds to the shift $\mathbf{p}\cdot\mathbf{v}\sub{d}$ in the shifted Fermi-Dirac $f(\mu + \mathbf{p}\cdot\mathbf{v}\sub{d})$ that is imposed at the source contact in the Boltzmann calculation, with the prefactor 0.5 for normalization. Fig.~\ref{fig:particle_trajectories}d compares the flow obtained using {\tt bolt}, which solves the Boltzmann equation, versus a particle simulation (fig.~\ref{fig:particle_trajectories}c). The currents are in good agreement, albeit with pronounced noise in the particle simulation.

\begin{figure*}[!htbp]
\begin{center}
\includegraphics[width=160mm]{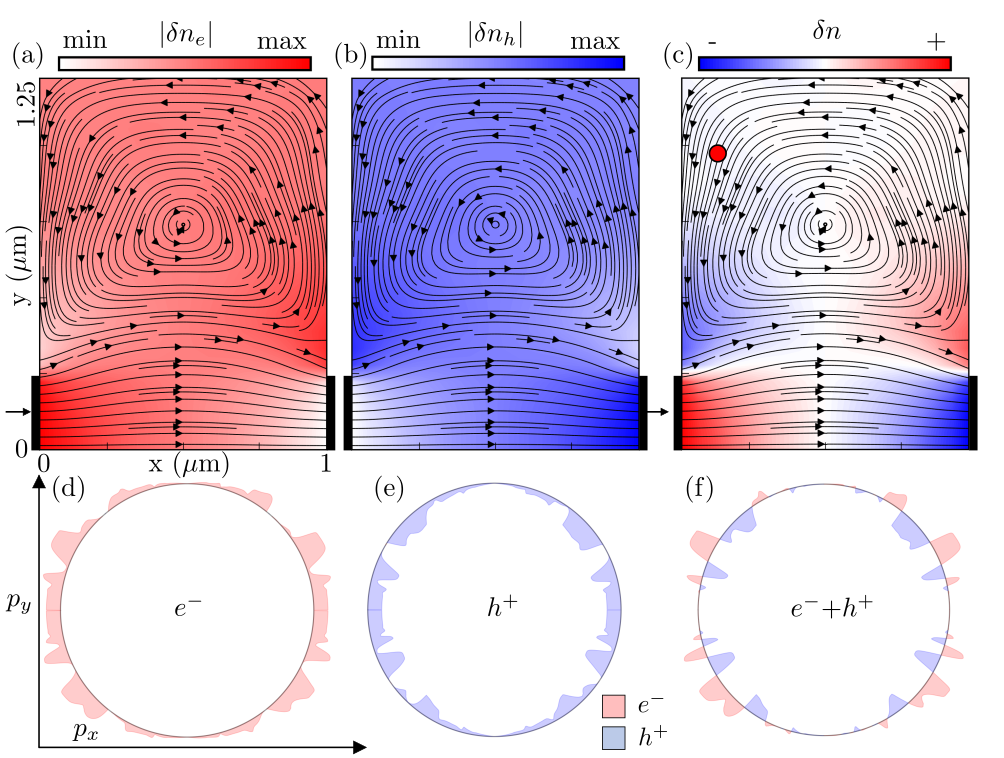}
\caption{(a,b) Current streamlines and particle density contours when only electrons (holes) are injected through the source (drain) contact. (c) Sum of the currents and densities in (a) and (b). (d-f) Corresponding local distributions, shown at the same spatial point as in fig.~\ref{fig:dist_func}(a-c). The distribution (f) exactly reproduces the local distribution presented in fig.~\ref{fig:dist_func}(b). The red and blue pockets in the distribution depict an excess of electrons and holes respectively.}\label{fig:classical}
\end{center}
\end{figure*}

Finally, note that the ballistic distribution shown in fig.~\ref{fig:dist_func}b of the main text is filled with particles \emph{and} holes. The presence of holes is distinctly non-classical, reminding us of the existence of the Fermi surface. Only injecting electrons through the source contact, as described above, cannot produce such a distribution. We \emph{also} need to inject holes through the drain contact. We show this explicitly by performing two separate {\tt bolt} simulations, one in which only electrons are injected through the source and another where only holes are injected through the drain. Fig.~\ref{fig:classical}(d,e) show the local distributions in each case. The correct ballistic distribution is obtained by the \emph{sum} (fig.~\ref{fig:classical}f) of the individual distributions.

\subsection{Hydrodynamic vs Ballistic vortices} \label{sec:hydro_ballistic}

Vortices are flow structures that are typical of fluids. How do the ballistic vortices compare to vortices in the hydrodynamic regime? To compute the vortical response in the hydrodynamic regime, it is easiest to resort to the fluid momentum conservation equation\cite{WhirlpoolsOrNot, levitov-2016, torre-2015},
\begin{align}
\nu \nabla^2 \mathbf{v} = \frac{e}{m} \mathbf{E}
\end{align}
where $\nu$ is the kinematic viscosity of the electron fluid. For a Fermi liquid, the viscosity is related to the scattering time scales, $\nu \simeq \tau\sub{mc}v\sub{F}^2/4$ \cite{principi-2016, bandurin-2016}. The hydrodynamic vorticity is then,
\begin{align}
\Omega\sub{hydro}(q\sub{y}) & = \frac{i e}{m} \frac{4}{\tau\sub{mc} q\sub{y} v\sub{F}^2} \label{eqn:vorticity_hydro}
\end{align}

The ballistic (\ref{eqn:vorticity_ballistic}) and hydrodynamic (\ref{eqn:vorticity_hydro}) vortical responses are very different. First, the hydrodynamic response depends on the interaction time scale $\tau\sub{mc}$. In Fermi liquids, $\tau\sub{mc} \sim T^{-2}$, and so the hydrodynamic vorticity vanishes as $T \rightarrow 0$. In contrast, the ballistic response is indeed finite even at $T=0$; as expected from a critical regime in which interactions are irrelevant.

Second, the ballistic response (\ref{eqn:vorticity_ballistic}) is the same for all spatial wavenumbers whereas the hydrodynamic response (\ref{eqn:vorticity_hydro}) only peaks for low wavenumbers. The wavenumber independent ballistic response implies that vortices at all scales are equally probable, leading to visually striking flow structures (fig.~\ref{fig:vortices}b). Such copious vortex generation in two-dimensions is highly non-classical and is in sharp contrast to the hydrodynamic response in which there is only one large vortex at the device scale (fig.~\ref{fig:vortices}c), even with very strong MC scattering ($\tau\sub{mc} = 0.01$ ps).

Finally, a crucial difference between ballistic and hydrodynamic vortices arises in the presence of disorder; the ballistic regime is \emph{far} more robust. While the vortices in the ballistic regime appear whenever $l\sub{mr}\gtrsim L\sub{x}/2$ (fig.~\ref{fig:conductivity}a-d), the requirement in the hydrodynamic regime is $l\sub{mr} \gtrsim (0.2/l\sub{mc}) L\sub{x}^2$\cite{WhirlpoolsOrNot}. We note here that the disorder requirements for vortex formation in the hydrodynamic regime are greatly mollified in AC transport, where they become equal to the ballistic regime \cite{chandra-2019}.

\newpage
\FloatBarrier

\begin{figure*}[!htbp]
\begin{center}
\includegraphics[width=160mm]{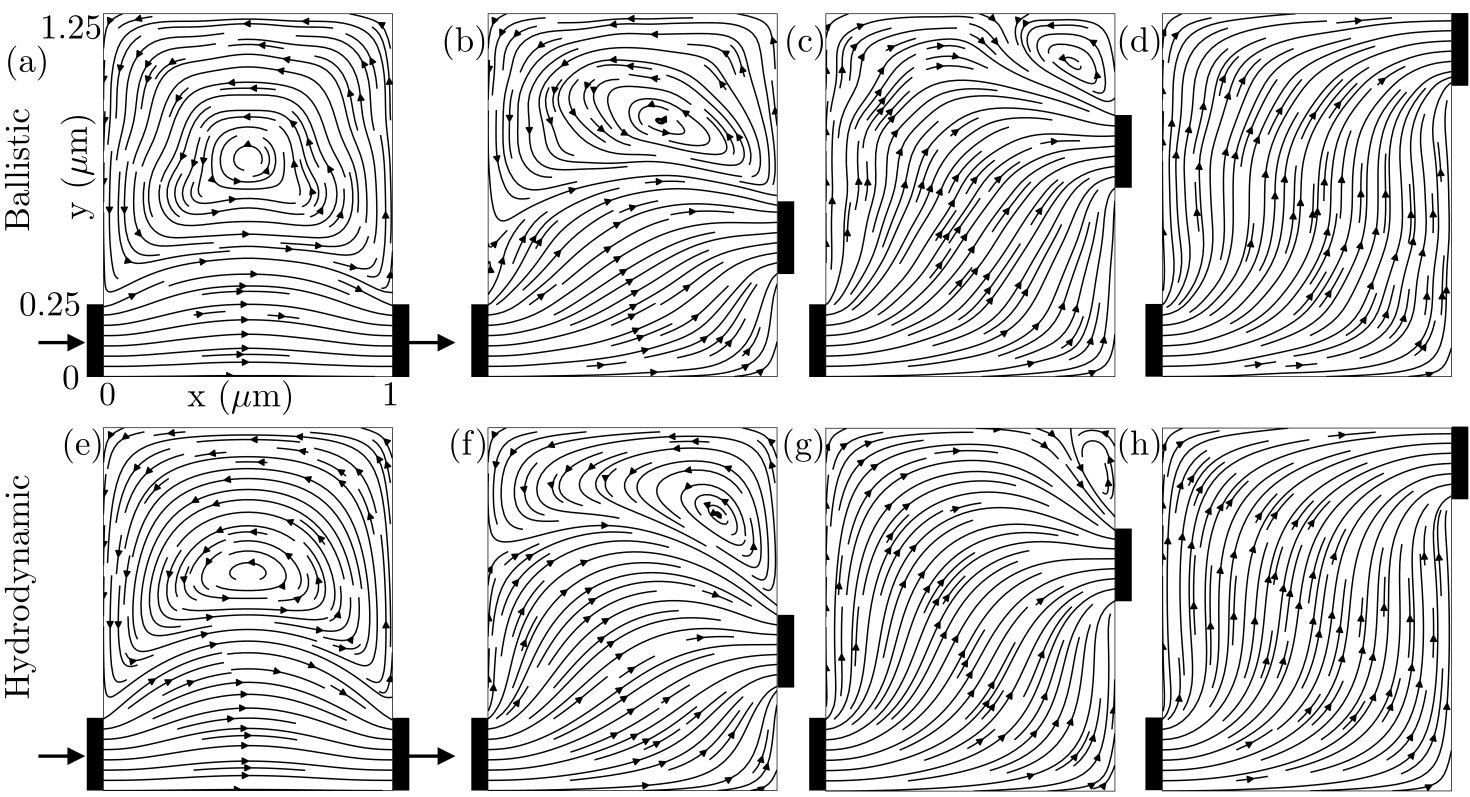}
\caption{Current streamlines in the ballistic (a-d) and hydrodynamic (e-h) regime for varying contact geometries. A regime with a finite vortical response, as in the ballistic and hydrodynamic cases, is a necessary condition for vortices but not sufficient. Vortices also require a conducive device and contact geometry; the diagonal contact configuration in (d, h) disallows vortex formation.}\label{fig:asymm_contact}
\end{center}
\end{figure*}

\begin{figure*}[!htbp]
\begin{center}
\includegraphics[width=160mm]{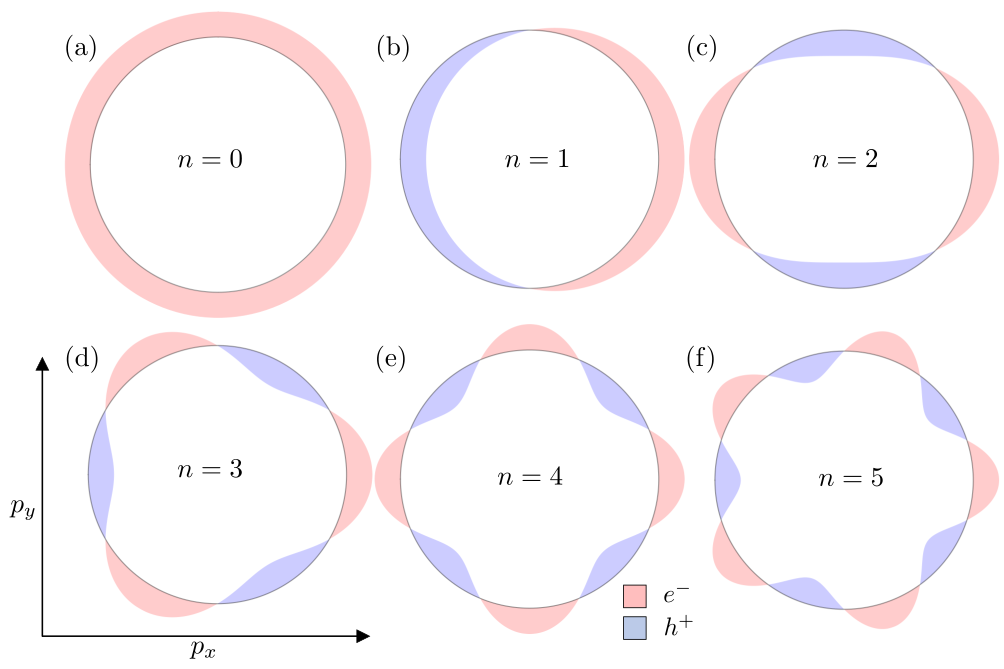}
\caption{Excitation modes of a circular Fermi surface. The black circle is the background equilibrium and the red (blue) regions are electron (hole) pockets. The $n=0$ mode, denoted by $A$ in fig.~\ref{fig:dist_func}, corresponds to a local change in the carrier density and is dominant in the Ohmic regime. The $n=1$ mode, denoted by $B$ in fig.~\ref{fig:dist_func}, corresponds to a shift and is the dominant mode in a hydrodynamic regime. The rest of the modes ($n \geq 2$) are fluctuations (shown here upto $n = 5$), and prevail in the ballistic regime. Note that all modes $n \geq 1$ are \emph{incompressible}; they have an equal number of electrons and holes, and cannot change the overall density. }\label{fig:modes_schematic}
\end{center}
\end{figure*}

\begin{figure*}[!htbp]
\begin{center}
\includegraphics[width=170mm]{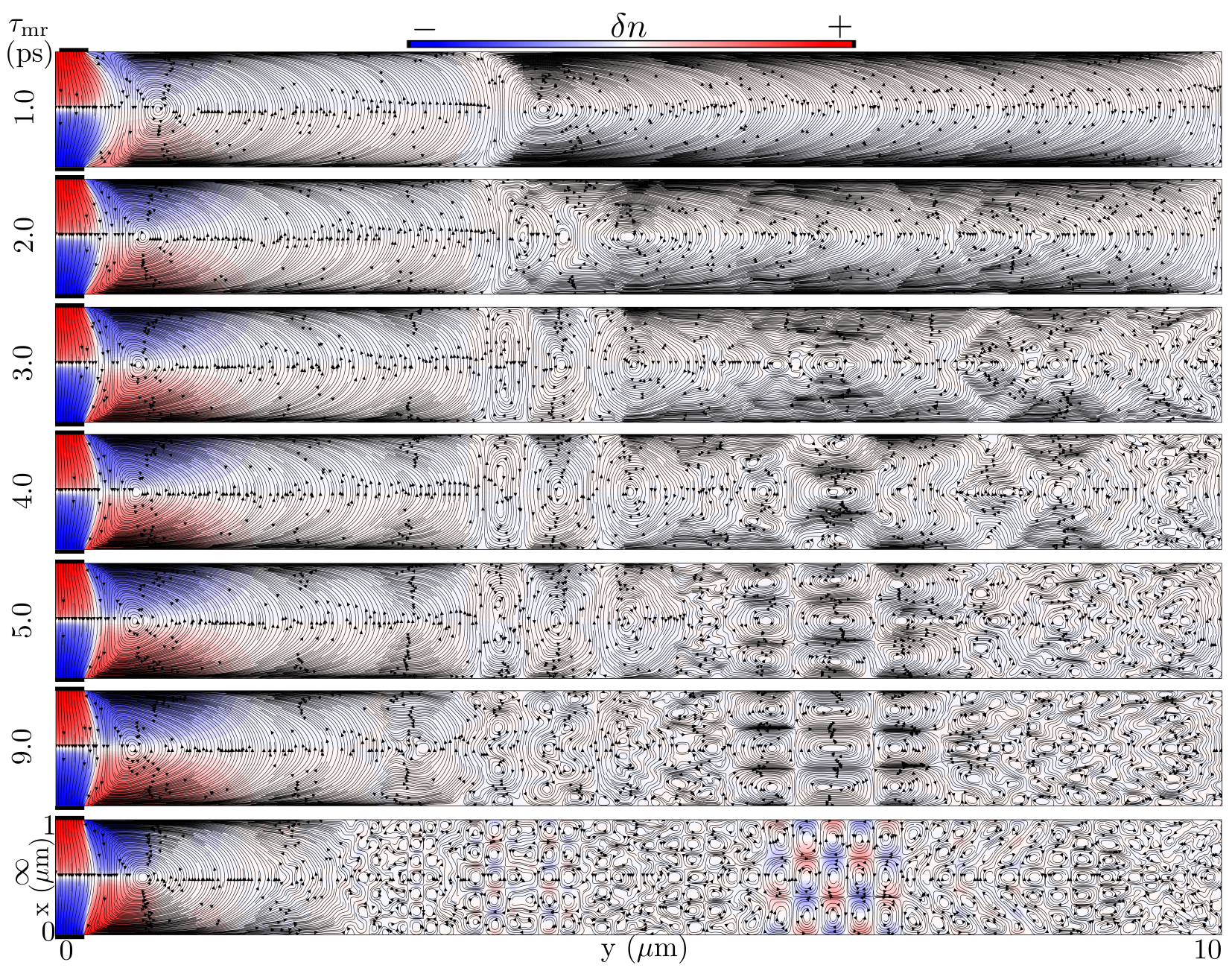}
\caption{The emergence of scale-invariant vortices in a wing with $\Gamma=9.75$. As the momentum-relaxing timescale $\tau_{mr}$ is progressively increased to go deeper into the ballistic regime, ever smaller vortices appear in the device. The flow in the $\tau\sub{mr}=\infty$ limit has structure on all scales, with tight correlation between the current vortices and the spatial density fluctuations.}\label{fig:1x10}
\end{center}
\end{figure*}

\begin{figure*}[!htbp]
\begin{center}
\includegraphics[width=170mm]{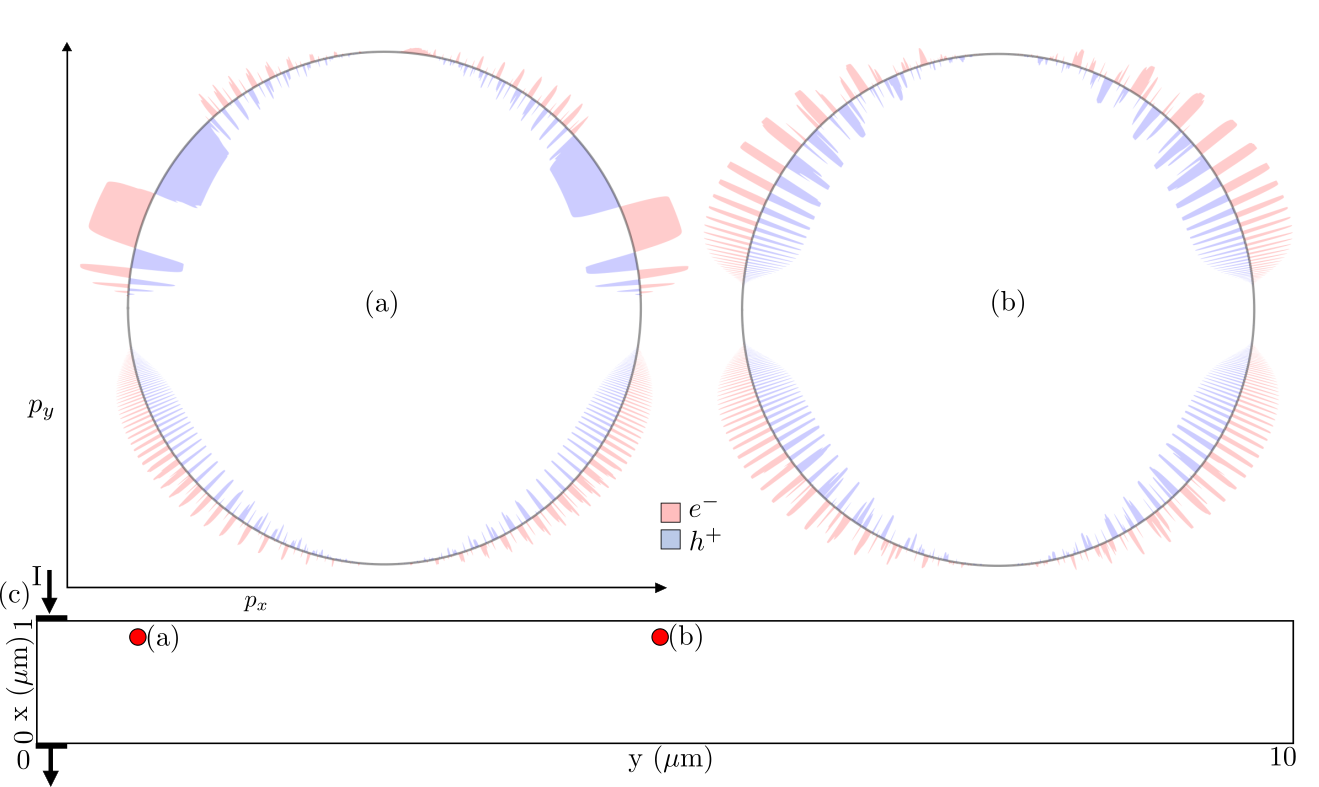}
\caption{(a,b) The local electron-hole fluctuations at different spatial locations (marked in the device schematic (c)) in a wing ($\Gamma=9.75$), deep inside the ballistic regime ($\tau_{mr} = \infty$). Comparing against the distribution in a wing with $\Gamma = 1$ (fig.~\ref{fig:dist_func}(b) of the main text), we see that devices with $\Gamma \gg 1$ have fluctuations extending to much finer angular scales (just as they do for spatial fluctuations).}\label{fig:fermi_surface}
\end{center}
\end{figure*}

\FloatBarrier

\begin{acknowledgments}
We thank Siddhardh Morampudi and Ravishankar Sundararaman for helpful discussions, and {\tt gpueater.com} for providing us with GPU (AMD Radeon) cloud instances. This research has been funded by {\it Quazar Technologies}.
\end{acknowledgments}

\end{document}